\documentstyle[preprint,prl,aps,epsfig]{revtex}
\tighten
\begin{document}
\bibliographystyle{revtex}
\draft
\title{Predicting total reaction cross sections for nucleon-nucleus
scattering}
\author{P.~K.~Deb$^1$, K.~Amos$^1$, S.~Karataglidis$^2$,
M. B. Chadwick$^2$, and D. G. Madland$^2$}
\address{$^1$School of Physics, University of Melbourne, Parkville,
Victoria, Australia, 3010 \\ $^2$Theoretical Division, Los Alamos National
Laboratory, Los Alamos, New Mexico, 87545}
\preprint{LA-UR-01-353}
\date{\today}
\maketitle
\begin{abstract}
Nucleon total reaction and neutron total cross sections to 300 MeV
for $^{12}$C and $^{208}$Pb, and for 65~MeV scattering spanning the
mass range, are predicted using coordinate space optical potentials
formed by full folding of effective nucleon-nucleon interactions with
realistic nuclear ground state densities. Good to excellent agreement
is found with existing data.
\end{abstract}
\pacs{}

In this letter we present the first predictions of integral
observables for both proton and neutron scattering for energies to
300~MeV (from $^{12}$C and $^{208}$Pb specifically) using, without
approximation, complex, nonlocal, coordinate-space optical potentials
formed by full folding realistic, effective, nucleon-nucleon ($NN$)
interactions with density matrices (hereafter called densities)
specified from credible models of the structure of the targets. That
predictions agree with observation forebodes well for the use of
such calculated values as are needed in a number of fields of study;
some of quite current interest.

An example of such a topical study concerns the transmutation of
long-lived radioactive waste into shorter-lived products , which
together with energy production, uses accelerator-driven systems
(ADS). These systems are being designed in the US, Europe, and Japan,
for providing an intense neutron source to a subcritical reactor. The
technology takes advantage of spallation reactions \cite{Wl00} in a
thick high-$Z$ target (such as Pb or Bi), where an intermediate-energy
proton beam induces nuclear reactions and the secondary nuclear
products, particularly lower-energy neutrons and protons \cite{Le99},
in turn induce further nuclear reactions in a cascade process. The
total reaction cross section plays a particularly important role since
the secondary particle production cross sections are directly
proportional to this quantity, which often are inputs to intranuclear
cascade simulations that guide ADS design.

In more basic science, the total reaction cross section is an
important ingredient to a number of problems in astrophysics, such as
nucleosynthesis in the early universe and for aspects of stellar
evolution. The density distribution of neutrons in nuclei is far less
well known than that of protons. By considering the integral observables of
both proton and neutron scattering from a given nucleus one may seek
direct information on the neutron root-mean-square radius; a property
sought in parity-violating electron scattering experiments \cite{Ba00}.

But most nucleon-nucleus ($NA$) total reaction cross sections have not
been measured. Thus we must have a reliable method for their
prediction. The usual vehicle for specifying $NA$ total reaction cross
sections has been the $NA$ optical potential; a potential most
commonly taken as a local parametrized function, usually of
Woods-Saxon type.  However, it has long been known that the optical
potential must be nonlocal and markedly so, although it has been
assumed also that the energy dependence of the customary
(phenomenological) model accounts for that.  Of more concern is that
the phenomenological approach is not truly predictive.  The parameter
values chosen, while they may be set from a global survey of data
analyses, are subject to uncertainties and ambiguities leading, for
example, to different values of the total reaction cross section while
yielding comparable elastic angular distributions.

We consider a predictive theory of $NA$ scattering to be one that is
{\em direct} for which all quantities required are defined {\em a
priori}, with no {\em a posteriori} adjustment of results.  With the
nucleus viewed as a system of $A$ nucleons, $NA$ scattering is
determined by an optical potential formed by a folding process. That
folding we choose to be of realistic interactions of the projectile
with each and every nucleon constituting the ground state target
density. Such microscopic approaches defining the $NA$ optical
potential have been quite successful in predicting angle-dependent
observables of elastic scattering \cite{Review}. In the coordinate
space approach, and for analyses that use the DWBA98 programs
\cite{Ra99}, the projectile-target nucleon interaction is not only
complex but also energy and density dependent.  We have used that
program herein to calculate all integral cross sections with realistic
effective interactions folded with reasonable descriptions of the
densities of nuclei to give credible optical potentials. Using those
optical potentials, differential cross sections and spin observables,
for proton scattering at many energies in the range 40 to 800~MeV (65
and 200~MeV in particular) and from diverse targets ranging from
$^3$He to $^{238}$U have been predicted and found to agree very well
with measured values.  The (nonlocal) optical potentials are complex
and energy dependent; properties that arise from mapping the effective
interactions to $NN$ $g$ matrices that are solutions of the
Brueckner-Bethe-Goldstone (BBG) equations for nuclear matter.  Details
of the specifications of the effective interactions, of the folding
process that gives the optical potential, and of the successful
predictions of differential cross sections and analyzing powers from
the scattering of protons at diverse energies and from diverse mass
targets, have been recently summarized \cite{Review}.

Formally, the nonlocal optical potentials can be written
\begin{eqnarray}
U(\vec{r}_1,\vec{r}_2; E) & = & \sum_n \zeta_n \left\{ \delta(\vec{r}_1
- \vec{r}_2) \int \varphi^{\ast}_n(\vec{s}) \, v_D( \vec{r}_{1s} ) \,
\varphi_n(\vec{s}) \; d^3s + \varphi^{\ast}_n(\vec{r}_1) \,
v_{Ex}(\vec{r}_{12}) \, \varphi_n(\vec{r}_2) \right\} \nonumber\\
& \Rightarrow & U_D(\vec{r}_1; E) + U_{Ex}(\vec{r}_1,\vec{r}_2; E)
\; ,
\label{NonSE}
\end{eqnarray}
where $v_D$ , $v_{Ex}$ are combinations of the components of the
effective $NN$ interactions, $\zeta_n$ are ground state one body
density matrices (which often reduce to bound state shell
occupancies), and $\varphi_n(\vec{r})$ are single nucleon bound
states.  All details and the prescription of solution of the
associated nonlocal Schr\"odinger equations are given in the recent
review~\cite{Review}.

The results to be discussed herein have been obtained by solving the actual
nonlocal Schr\"odinger equations defined with potentials as given by
Eq.~(\ref{NonSE}). For the present calculations, the effective $NN$
interactions have been defined by their mapping to the BBG $g$
matrices of the BonnB $NN$ interaction \cite{Ma89}. When those
effective interactions are folded with the appropriate ground state
target densities we obtain the ``g-folding'' optical
potentials. Calculations have also been made using solely the free
$NN$ $t$ matrix in this prescription, the results of which are
``$t$-folding'' optical potentials.

The specification of the nuclear ground state density is taken from a
given nucleon-based model of structure. For the light nuclei ($A \leq
16$) we use complete (no core) multi-$\hbar\omega$ shell models
\cite{Review}. For $^{12}$C, specifically, we have constructed a
complete $(0+2)\hbar\omega$ shell model using the WBT interaction of
Warburton and Brown \cite{Wa92}. For $^{208}$Pb, the other nucleus
under investigation, we have used both a simple packed shell model with
an oscillator specified by an $A^{1/3}$ rule and a Skyrme-Hartree-Fock
(SHF) specification \cite{Ba00} of the ground state.

The first objective of specifying the optical potentials is to define
the $S$ matrix, or equivalently the (complex) phase shifts
$\delta^{\pm}_l(k)$ which relate by
\begin{equation}
S^{\pm}_l \equiv S^{\pm}_l(k) = e^{2i\delta^{\pm}_l(k)} =
\eta^{\pm}_l(k)e^{2i\Re\left[ \delta^{\pm}_l(k) \right] }
\end{equation}
where
\begin{equation}
\eta^{\pm}_l \equiv \eta^{\pm}_l(k) = \left| S^{\pm}_l(k) \right| =
e^{-2\Im\left[ \delta^{\pm}_l(k) \right] } \;.
\end{equation}
In terms of these and with $E \propto k^2$, the elastic and total reaction
(absorption) cross sections are given by
\begin{eqnarray}
\sigma_{\mbox{el}}(E) & = & \frac{\pi}{k^2} \sum^{\infty}_{l=0}
\left\{ (l+1) \left| S^+_l -1 \right|^2 + l \left| S^-_l -1
\right|^2 \right\} \;, \nonumber \\
\sigma_R(E) & = & \frac{\pi}{k^2} \sum^{\infty}_{l=0} \left\{ (l+1)
\left[ 1 - \left( \eta^+_l \right)^2 \right] + l \left[ 1 - \left(
\eta^-_l \right)^2 \right] \right\} \;,
\end{eqnarray}
respectively. The total cross section, $\sigma_{\mbox{\tiny TOT}}(E)$,
is the sum of these two cross sections.

\begin{figure}
\centering\epsfig{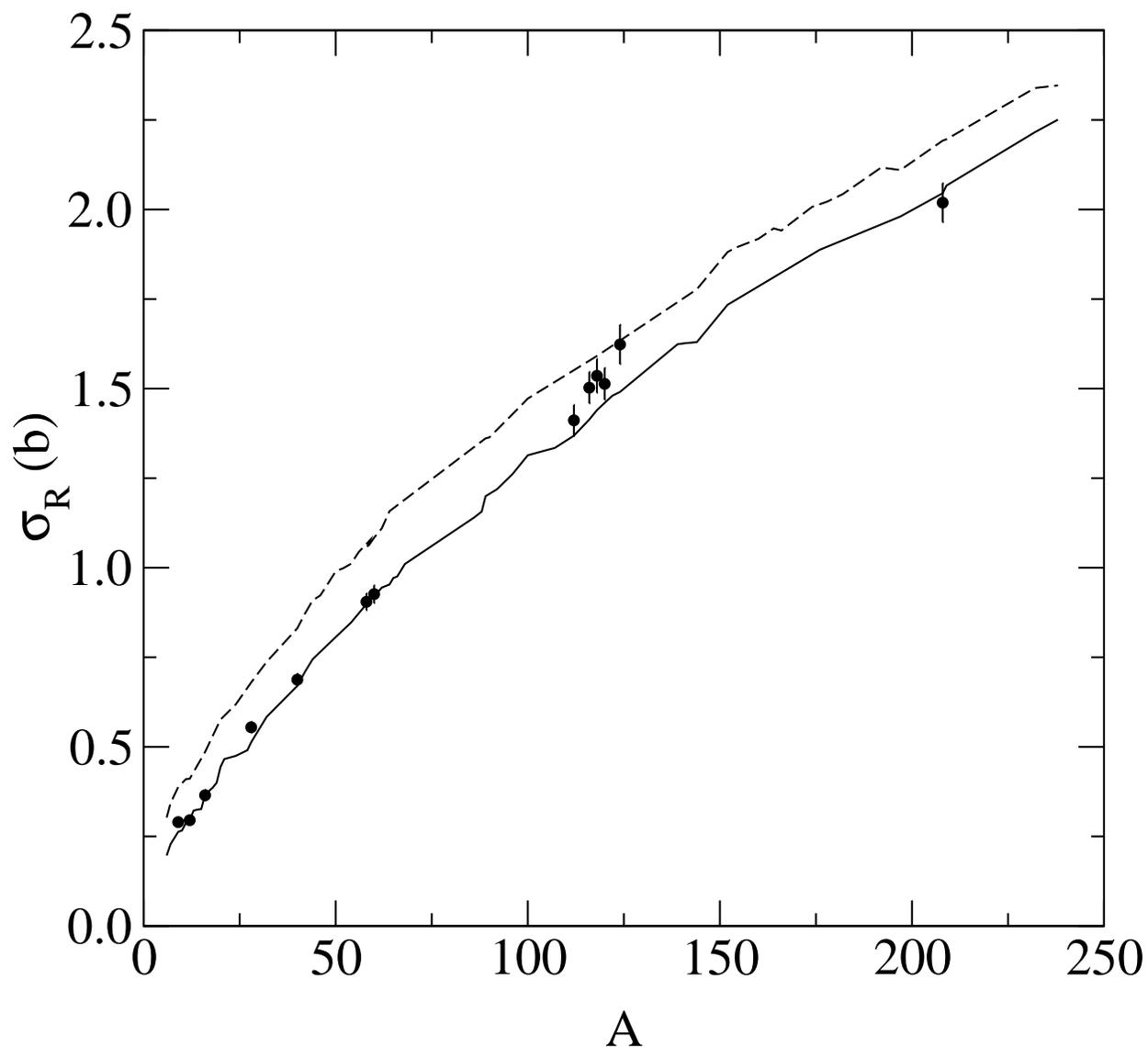}
\caption[]{Total reaction cross section at 65.5~MeV as a function of
target mass. The $g$-folding (solid line) and $t$-folding (dashed
line) results are compared to the data of Ingemarsson \cite{In99}.}
\label{mass}
\end{figure}

In Fig.~\ref{mass}, proton total reaction cross-section data at
65.5~MeV \cite{In99} from a diverse set of nuclei are compared with
our predictions obtained using both the $g$ and $t$ folding optical
potentials. Quite clearly, with large complete space shell model
densities \cite{Review} for masses 9 to 12, complete $0\hbar\omega$
densities for $sd$-shell nuclei to $^{40}$Ca, packed structure models
for the Ni and Sn isotopes, and the SHF model for $^{208}$Pb, the
medium modifications generating the $g$ folding potentials are
essential to give agreement with experiment.

There are a number of small breaks in the mass variation of the $g$
folding results from a smooth trend. These reflect density variations
with isotopes created even by the simple $0\hbar\omega$ model and
which influence how specifics of the effective interactions are
selected in folding. Such variations have lesser influence in defining
the $t$-folding optical potentials; hence the smoother variation with
mass of the $t$-folding total reaction cross sections. However, the
$t$-folding results consistently overestimate the data.

\begin{figure}
\centering\epsfig{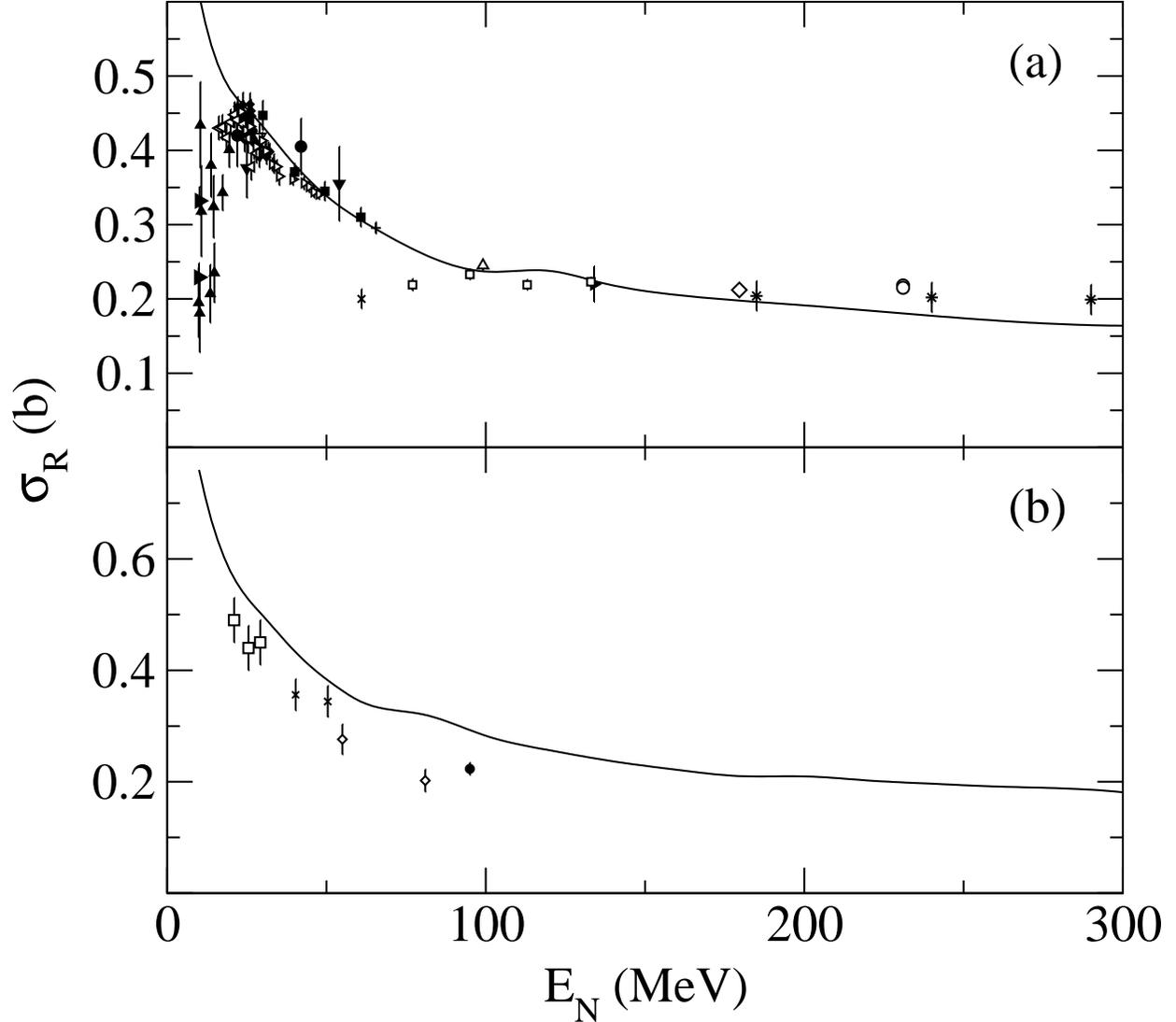}
\caption[]{Energy dependence of the total reaction cross section for
the scattering of nucleons from $^{12}$C. In (a), the results for the
scattering of protons are compared with data taken from many sources
as listed in the compilation by Carlson \cite{Ca96}, and from more
recent data of Ingemarsson {\em et al.} \cite{In99}. In (b), the
results for the scattering of neutrons are compared with the data of
Refs. \cite{Za81} (crosses), \cite{Mc58} (squares), \cite{Vo56}
(diamonds), and \cite{De50} (circle).}
\label{c12reac}
\end{figure}

The total reaction cross sections for proton and neutron scattering
from $^{12}$C up to 300~MeV are presented in sections (a) and (b),
respectively, of Fig.~\ref{c12reac}. The proton data have been
obtained from many experiments \cite{In99,Ca96}. The neutron data were
obtained from References \cite{Za81,Mc58,Vo56,De50}.  The solid curves
are our $g$-folding model predictions. Clearly, with the exception of
the data below 20 MeV and of two data points at 61 and 77~MeV, our
predictions of the proton total reaction cross sections are quite good.  The
mismatch below 20 MeV does not vitiate our method of analysis for
higher energies. The spectrum of $^{12}$C, and also of the compound
mass-13 system, has many discrete states and so nucleon scattering
will reflect effects of (discrete) thresholds, isolated compound
nuclear resonances, etc. Our first order folding model of potential
scattering has not been designed to accommodate such specific
structures and processes. The exceptional data points also should be
discounted. Not only do they differ markedly from the trend of other
data in the vicinity but also there are data points at 60.8 and
65.5~MeV that conflict with them.  The neutron total reaction
cross-section data vary with energy as do our $g$ folding predictions
but we overestimate the extracted experimental values by some 10-20\%.
However, the neutron total reaction cross section data are difficult
to measure and further judgement of these results should be held in
abeyance until we consider the neutron total cross sections.

\begin{figure}
\centering\epsfig{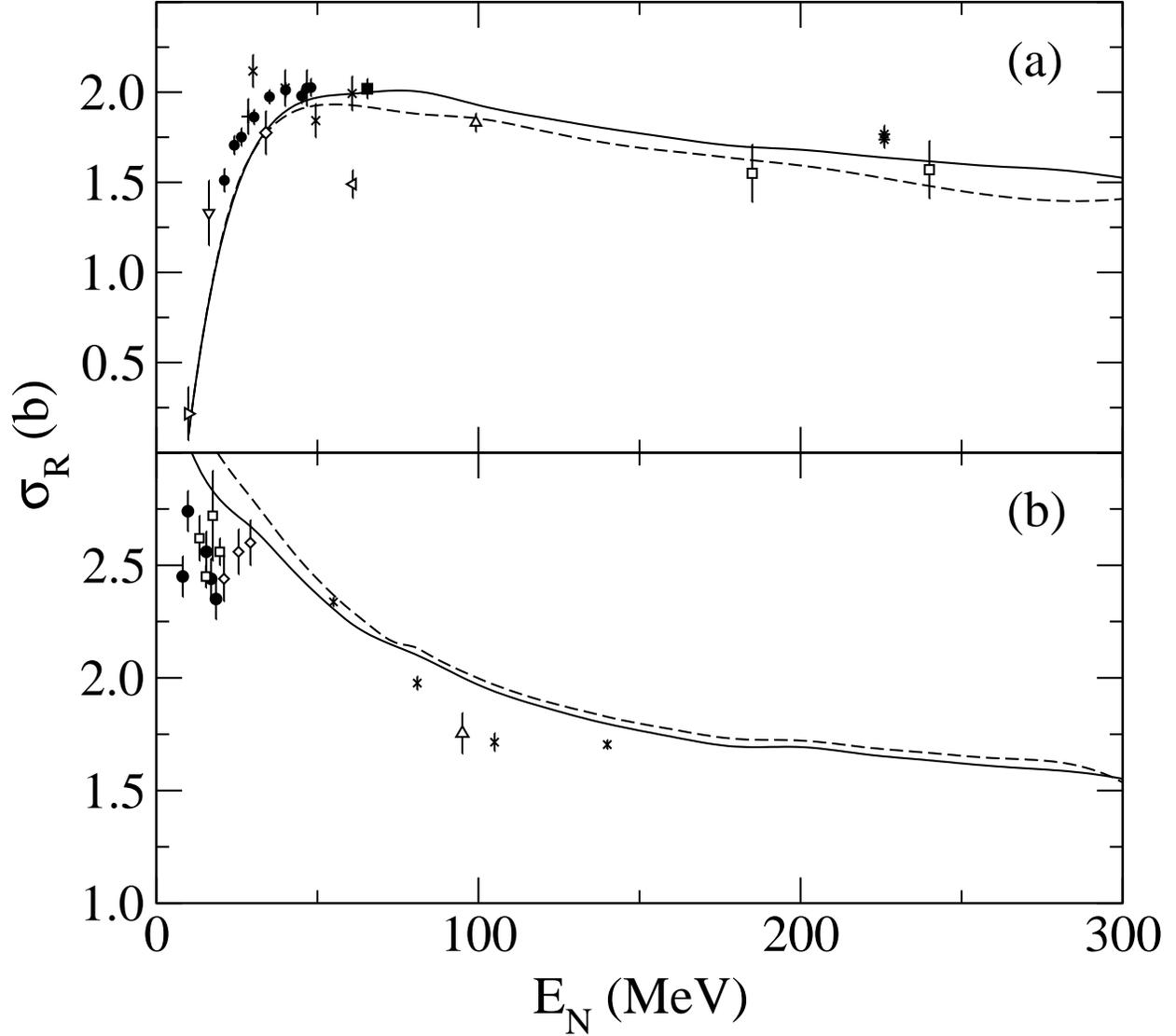}
\caption[]{As for Fig.~\ref{c12reac} but for the scattering of
nucleons from $^{208}$Pb. The data for the neutron total reaction
cross sections were from Refs. \cite{Mc58} (diamonds), \cite{Vo56}
(crosses), \cite{De50} (triangle), \cite{Bo59} (circles), and
\cite{De62} (squares). The additional dashed curves were obtained
using simple $0\hbar\omega$ shell model densities.}
\label{pbreac}
\end{figure}

The total reaction cross sections for proton and neutron scattering
from $^{208}$Pb are displayed in Figs.~\ref{pbreac}(a) and
\ref{pbreac}(b), respectively.  The proton data were taken from
references \cite{In99} and \cite{Ca96} and the neutron data from
references \cite{Mc58,Vo56,De50,Bo59,De62}.  We compare those data
again with the predictions obtained using two sets of $g$ folding
optical potentials.  In all cases, at each energy, the same effective
interaction appropriate for that energy has been used to give the
optical potentials.  The dashed curves are predictions one obtains
when simple $0\hbar\omega$ shell model densities are used in the
folding process, while the solid curves depict our predictions found
by folding with the SHF densities. The differences between these
predictions are not large, but in comparison with the data, the SHF
model is preferred.  Both calculations slightly underestimate the data
in the range of energies 10 to 40~MeV.  In contrast, our results
overestimate most of the neutron total reaction cross section data.
But as with the $^{12}$C case, any judgement of the neutron data
mismatch should await consideration of the total scattering cross
sections.

However, note that proton (neutron) scattering from nuclei are most
influenced by the folding with the neutron (proton) densities given
the preponderance of the isoscalar $^3S_1$ $NN$ interaction in such
foldings.  Also, with increasing energy, the central attributes of the
optical potentials gain in importance in determining the scattering
phase shifts and hence the total reaction cross sections.  At low
energies, not only do relatively few partial waves contribute
to the calculations, but also the centrifugal barrier more effectively
screens the waves from the influence of the bulk of the optical potential,
hence lower energy scattering tends to be a surface effect.  Thus, as
the SHF and $0\hbar\omega$ model densities differ most noticeably in
the nuclear interior, the proton scattering cross sections found with
these two model densities differ by about 100~mb for energies above
100~MeV but are more equivalent at lower energies.  The discrepancy
with data at lower energies then indicates that the surface
distributions of neutrons in $^{208}$Pb are even more extended than
given by either model.

\begin{figure}
\centering\epsfig{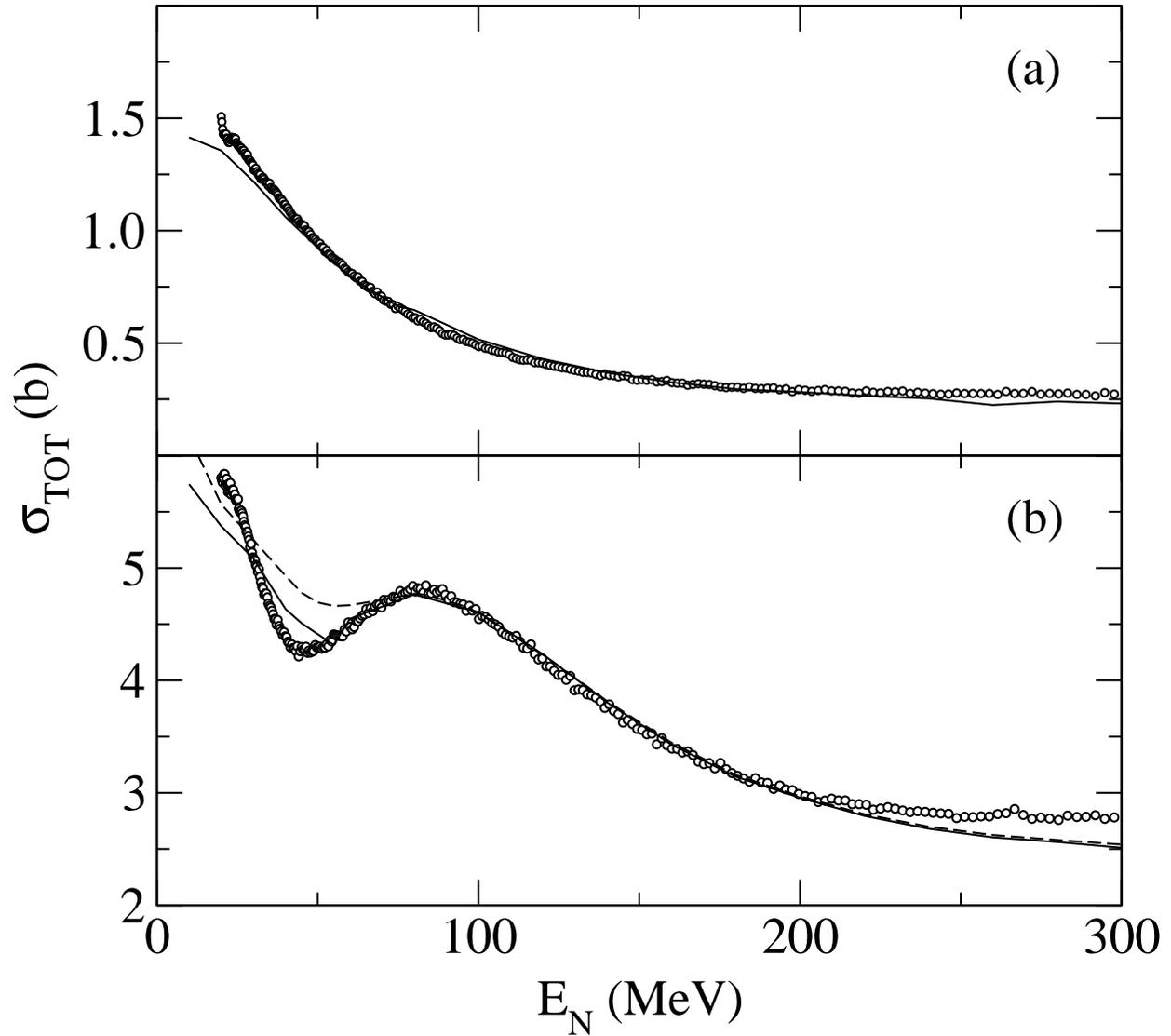}
\caption[]{Total cross section for the scattering of neutrons from
$^{12}$C (a) and $^{208}$Pb (b). The data of Finlay {\em et al.}
\cite{Fi93} are compared to the results of the $g$ folding
calculations. The solid and dashed curves are as for
Fig.~\ref{pbreac}.}
\label{total}
\end{figure}

The total scattering cross sections of neutrons from $^{12}$C and
$^{208}$Pb are displayed in Figs.~\ref{total}(a) and \ref{total}(b),
respectively.  The data shown therein are those of Finlay {\em et al.}
\cite{Fi93} and they are compared with our $g$ folding optical
potential predictions. Overall, our predictions are in good agreement
with the data and the SHF result in $^{208}$Pb gives a better
reproduction of the Ramsauer diffraction structure in the 30 to
200~MeV region in comparison to the recent semi-microscopic optical
model result \cite{Ba98}. Noticeably, for both nuclei, there is no
overestimation of the data in the 50 to 100 MeV region in contrast to
the total reaction cross-section comparisons. The $g$ folding optical
potentials lead also to excellent reproductions of differential
cross-section data \cite{Review} with implications that the total
elastic cross sections are well predicted. Thus for neutron
scattering, we are confident that our total reaction cross sections
are also realistic.

We have demonstrated that our $g$ folding optical potential procedure
gives integral observables in good agreement with existing
data. Particularly, we conclude that the low energy data (typically to
60~MeV) are sensitive to the surface density properties and of the
neutrons for the heavy nuclei in particular.  At higher energies, the
surface densities remain important, but it is the bulk density that
largely determines the scattering. As a consequence, an analysis of
proton and neutron integral cross sections with energy can shed light
on the neutron density distribution in nuclei.

This work was supported by a grant from the Australian Research
Council and also by DOE Contract no. W-7405-ENG-36.
\bibliography{Reaction}

\end{document}